\documentclass[prl,aps,twocolumn,groupedaddress,floats,showpacs,final]{revtex4-1}
\usepackage{graphicx}
\usepackage{dcolumn}
\usepackage{bm}
\usepackage{color}
\definecolor{blue}{rgb}{0.3,0.3,0.9}

\begin{document}

\title{Excitation spectrum of a supersolid}

\author {S. Saccani$^1$, S. Moroni$^{1}$ and M. Boninsegni$^2$}

\affiliation {$^1$SISSA Scuola Internazionale Superiore di Studi Avanzati and
        DEMOCRITOS National Simulation Center,
          Istituto Officina dei Materiali del CNR
         Via Bonomea 265, I-34136, Trieste, Italy
}
\affiliation {$^2$Department of Physics, University of Alberta, Edmonton, Alberta, Canada T6G 2G7}
\date{\today}
\begin{abstract}

Conclusive experimental evidence of a supersolid phase in any known condensed matter system is presently lacking.
On the other hand, a supersolid phase has been recently predicted for a system of spinless bosons in continuous space, interacting via a broad class of soft-core, repulsive potentials. Such an interaction can be engineered in assemblies of ultracold atoms, providing a well-defined pathway to the unambiguous  observation of this fascinating phase of matter. 
In this article, we study by first principle computer simulations  the elementary excitation spectrum of  the supersolid, and show that it features two distinct modes, namely a solid-like phonon and a softer collective excitation, related to broken translation and gauge symmetry respectively.

\end{abstract}



\maketitle
The existence, physical properties and possible observation of a supersolid phase of matter, have been the subject of intense scientific debate for decades\cite{Gross,Andreev_Lifshitz,Thouless,Leggett,Chester}. Such a phase should feature the seemingly antithetic properties of crystalline order and superfluidity. The bulk of the investigative effort, aimed at identifying and characterizing this phase in known condensed matter systems, has focused on the most obvious candidate, namely solid $^4$He. Presently, however, consensus is lacking as to whether the existing experimental evidence unambiguously points to supersolidity \cite{rmp}.
\\ \indent
In recent times, attention has turned to the field of ultracold atoms as a possible experimental venue affording supersolid behaviour. Computer simulations have yielded evidence of a supersolid phase in a two-dimensional system of spinless bosons in continuous space (i.e., {\it not} on a lattice), interacting through a specific class of soft-core repulsive potentials\cite{cinti,saccani}. At low temperature and high density, despite the repulsive character of the interaction, particles pile up in clusters, in turn forming a periodic array, to which we refer as ``cluster crystal"\cite{Likos1}: this is a lattice with a basis of $K$ particles of the same kind and null basis vector. This solid turns superfluid at temperatures sufficiently low that phase coherence can be established through quantum hopping of particles across lattice sites (i.e., adjacent clusters). 
\\ \indent
To our knowledge, the particular soft-core interaction that underlies this supersolid phase does not occur in ordinary condensed matter systems; however, it can be engineered for ultracold atoms\cite{lukin}. This could pave the way for a clear-cut experimental realization of this up to now elusive phase. Thus, soft-core bosons may be regarded as the archetypal supersolid, an ideal playground wherein fundamental properties of the supersolid phase of matter can be explored. Of particular interest is the spectrum of elementary excitations of a supersolid, which offers access to arguably even more cogent information on the physics of a system, than structural or energetic properties of the ground state. 
\\
\indent
There are a number of outstanding issues, regarding the basic features of the dynamic response function of a supersolid. For example, it is not obvious how the excitation spectrum would combine specific traits of a solid and of a superfluid, i.e.,  whether two separate Goldstone modes should be present, reflecting the two broken symmetries, or a single mode of distinct, different character. Also of  interest is establishing whether the excitation spectrum of a superfluid system, that also breaks translational invariance, displays a roton minimum. These fundamental issues go to the heart not only of supersolidity, but of our current, general understanding of quantum many-body systems. It is also worth noting that the experimental study of the dynamic structure factor in assemblies of ultracold atoms has recently begun\cite{Stamper-Kurn,inguscio}.
\\ \indent
In this Article, we report results of a calculation of the low-temperature excitation spectrum of density fluctuations\cite{sqom} in the dense fluid, cluster crystal and supersolid cluster crystal phases. We take the soft-disk (SD) potential as representative\cite{saccani} of the class of interactions described above, which support the phases of interest here. 
\\ \indent 
We find that a new branch of acoustic modes appears, not specifically related to the phonon-maxon-roton of the superfluid or the phonons in the solid. This branch, which has an analogue in the spectrum of the Bose-Hubbard model (BHM) in its superfluid phase\cite{capogrosso}, can be related to the breaking of the gauge symmetry\cite{menotti,Yoo}.

\begin{figure}
\begin{center}

\includegraphics[scale=0.6,angle=-90]{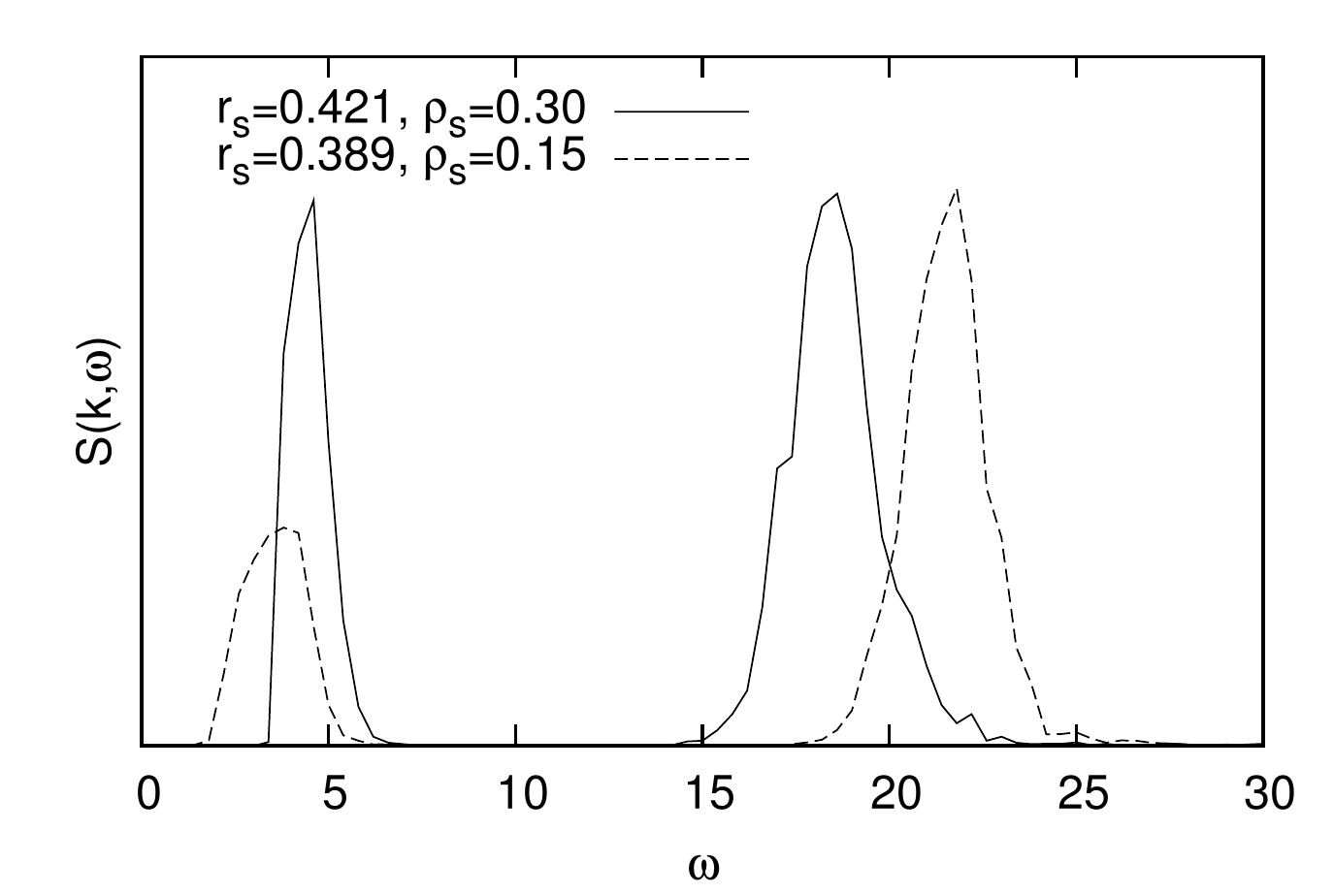}

\caption{\label{spectrum_confronto} {\bf Comparison of $S(k,\omega)$ at different densities.} They are calculated at the edge of the brillouin zone, for $k= \frac{2\pi}{d\sqrt{3}}$. The solid (dashed) line pertains to a supersolid with $r_s=0.421$ and $\rho_s=0.30$ ($r_s=0.398$ and $\rho_s=0.15$).
}
\end{center}
\end{figure}

The Hamiltonian of the system in reduced units is
\begin{eqnarray}\label{mod}
\mathcal{H}=-\frac{1}{2}\sum_{i=1}^N \mathbf{\bigtriangledown}_{i}^2+ D \sum_{i>j} \Theta(1-r_{ij}),
\end{eqnarray}
where $r_{ij}$ is the distance between particles $i$ and $j$ and $N$ is the particle number. The diameter $a$ of the soft disks is taken as the unit of length, while $\epsilon_\circ = {\hbar^2}/{ma^2}$ is the energy unit. The system is enclosed in a cell of area $A$, with periodic boundary conditions. We  express the density $\rho$ through the dimensionless parameter $r_s=1/\sqrt{\rho a^2}$, i.e., the mean interparticle distance.
We focus on a range of density where the mean number $K$ of particles per site in the cluster crystal phase is relatively high (roughly between ten and twenty)\cite{note}.\\ \indent
We use the Worm Algorithm in the continuous-space path-integral representation\cite{worm,worm2} to simulate the system described by (\ref{mod}) in the grand canonical ensemble (i.e., at fixed temperature $T$, area $A$ and chemical potential $\mu$). The simulation gives an unbiased, accurate numerical estimate of the imaginary-time intermediate scattering function
\begin{eqnarray}\label{Fqt}
F(\mathbf{k},\tau) =\langle{\hat\rho}_{\mathbf{k}}(\tau) {\hat\rho}^\dagger_{\mathbf{k}}(0)\rangle/N,
\end{eqnarray}
where ${\hat\rho}_\mathbf{k} = \sum_j e^{i \mathbf{k}  \cdot {\mathbf{r}_j}} $ is the density fluctuation operator at wavevector $\mathbf{k}$ and the brackets denote a thermal average.
The dynamic structure factor $S(\mathbf{k},\omega)$, which measures the excitation spectrum of the density fluctuations, is related to $F(\mathbf{k},\tau)$ 
via an inverse Laplace transform:
\begin{eqnarray}\label{sqomega}
F(\mathbf{k},\tau) =\int \mathrm{d}\omega e^{-\tau\omega} S(\mathbf{k},\omega).
\end{eqnarray}
It is well known that there exists no general scheme to invert a Laplace transform from noisy data, in a  way that is reliable, accurate and controlled. However, for physical spectra whose dominant contribution is given by  a few well-defined peaks,  some techniques are able to identify satisfactorily locations and spectral weights of those peaks. In this work, we made use of  the Genetic Inversion via Falsification of Theories (GIFT)\cite{Vitali} approach for the numerical inversion of the Laplace transform (\ref{sqomega}).
When applied to superfluid $^4$He, the GIFT method has been shown to separate correctly the sharp quasiparticle peak of the phonon-maxon-roton elementary excitation from the broad multiphonon contribution\cite{Vitali}, whereas the more commonly adopted Maximum Entropy scheme\cite{maxent} tends to merge both structures\cite{sqom}. Alongside with this method, using the information on the number $n$ of excitations visible in the reconstructed spectrum, we compute the energy of the $n$ observed excitations by assuming the spectrum $S(\mathbf{k},\omega)$ as a function of $\omega$ only is constituted by $n$ delta functions ($n$-pole approximation) and fit their positions and strengths to the available $F(\mathbf{k},\tau)$ data.
\\ \indent
\begin{figure}
\begin{center}

\includegraphics[scale=0.55,angle=-90]{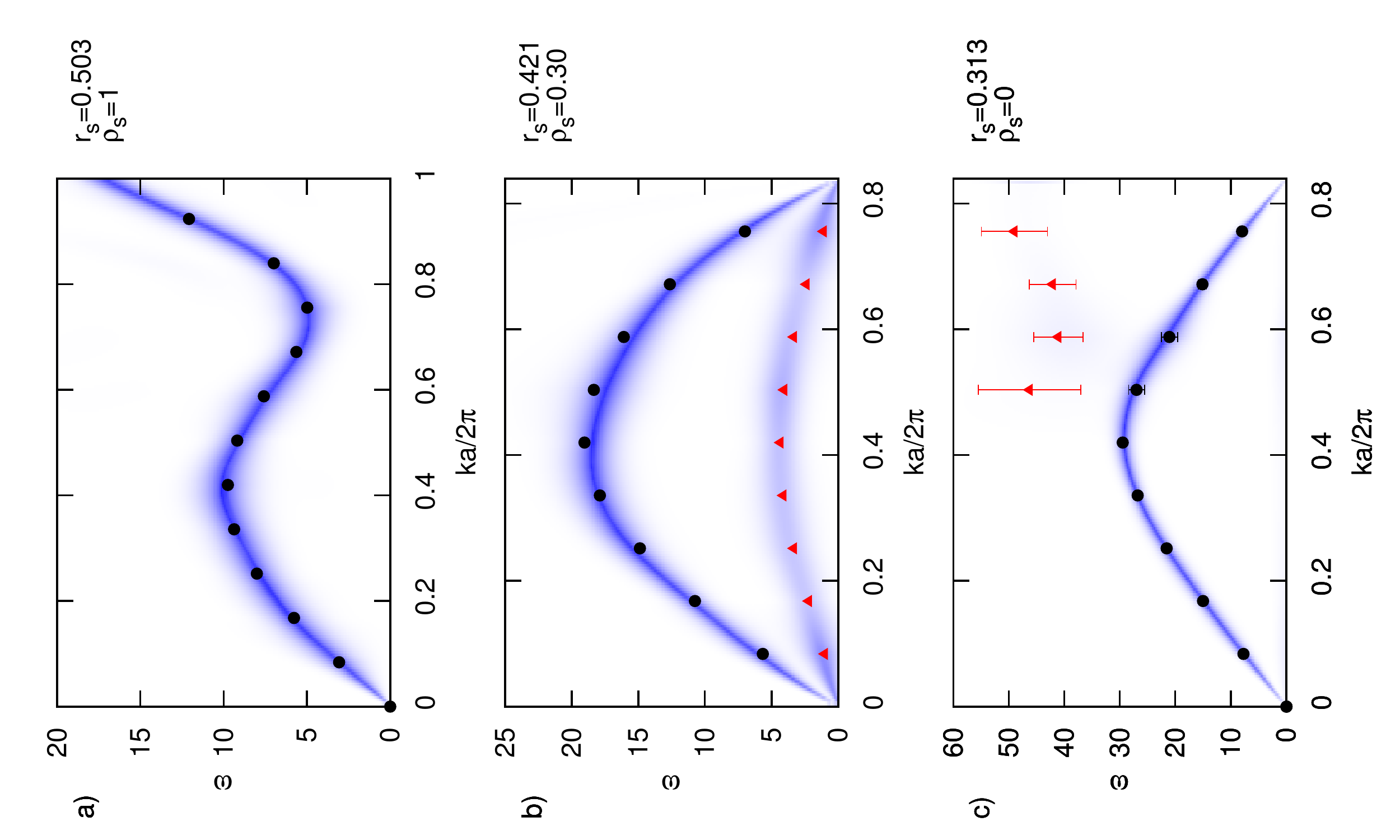}

\caption{\label{sqom_all} {\bf Dynamic structure factor $S(\mathbf{k},\omega)$ in various phases.} Panels refer to a) a superfluid, b) a supersolid, and c) a non-superfluid cluster crystal. The colormap is obtained by smoothing and interpolation of the calculated GIFT spectra. In order to emphasize the dispersion of the spectrum, for each $\mathbf{k}$ the dynamic structure factor is rescaled to a common maximum value in all panels. The datapoints are obtained from the n-pole approximation (see text); when errorbars are not reported, these are of the order of the symbol size or smaller. For the modulated phases b) and c), the primitive vectors of the Bravais lattice are $d(1,0)$ and $d(1/2,\sqrt{3}/2)$ with $d=1.375a$; the wavevector spans the range $0$--$\frac{4\pi}{d\sqrt{3}}$ along the direction $[0,1]$. The mean site occupation is $K=9.2$ in b) and $K=16.7$ in c). Also given are the values of $r_s$ (mean interparticle distance) and $\rho_s$ (superfluid fraction).} 
\end{center}
\end{figure}
Figure \ref{sqom_all}, colormap, shows GIFT reconstructions of the dynamical structure factor $S(\mathbf{k},\omega)$ in the superfluid, supersolid and cluster crystal phases; datapoints are obtained from the $n$-pole approximation instead.
In the superfluid phase (Fig. \ref{sqom_all} a), the spectrum is characterized by the usual phonon-maxon-roton dispersion, with the notable peculiarity that the roton minimum is located just short of $2 \pi /a$, rather than around $2 \pi /r_s$. This suggests that the incipient crystallization takes place with a lattice parameter larger than the mean interparticle distance. Indeed, upon increasing $\mu$, the superfluid undergoes a first order phase transition into a triangular cluster (super)solid with a lattice spacing $d$ somewhat larger than $a$\cite{saccani}.
\\ \indent
The spectrum of the cluster crystal is also  standard (we have studied longitudinal excitations only). Figure \ref{sqom_all}c) shows that, within the first Brillouin zone, most of the spectral weight is concentrated in an acoustic phonon band.
We observe a non-negligible zero-frequency contribution at all wavelengths, representing a diffusive mode of lattice defects (phase-incoherent hopping of particles between multiply occupied sites). We also observe for some $\mathbf{k}$-points the presence of an optical mode, which might be the analog of the highly degenerate breathing mode of individual clusters observed for classical cluster systems\cite{Neuhaus}, but we do not determine its dispersion relation here. 
\\ \indent
The excitation spectrum of the supersolid phase (Fig. \ref{sqom_all}b), is the central result of this work. The spectral weight is clearly partitioned in two distinct branches. The higher-energy mode is a longitudinal acoustic phonon, with a linear dispersion at small $\mathbf{k}$ and near the reciprocal lattice vector (the end of the $\mathbf{k}$ scale in the figure), and frequencies between the phonon-maxon of the liquid and the longitudinal phonon of the solid. 
This assignment is further supported by the following analysis. We computed the average potential $v_{\rm test}(r)$ felt by a test particle across a lattice site  (upper panel of Figure \ref{density_confronto}, also showing the particle density profile). We define a force constant by fitting a quadratic potential to the bottom of $v_{\rm test}(r)$, and obtain the phonon frequency of a harmonic crystal with that force constant and the average mass of a cluster, given by $K$. The sound velocity of the harmonic crystal is in satisfactory agreement (to within $\sim 10\%$) with the slope of the phononic branch of the supersolid displayed in Figure \ref{sqom_all} b).
\\ \indent
The lower branch of the supersolid spectrum is also acoustic.
\begin{figure}
\begin{center}

\includegraphics[scale=0.55,angle=-90]{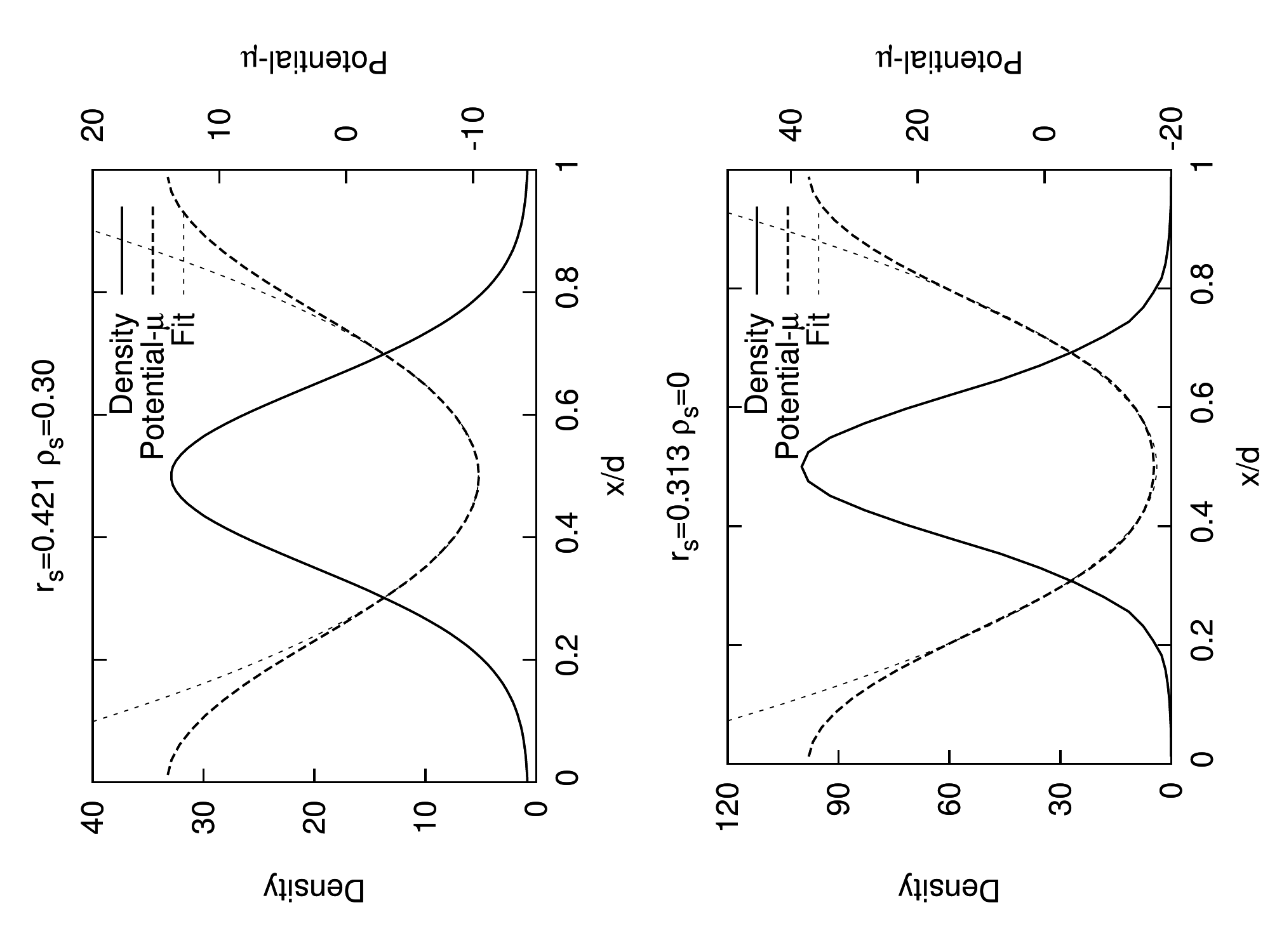}

\caption{\label{density_confronto} {\bf Density profile and potential felt by a test particle between two adjacent sites}. The upper (lower) panel corresponds to the supersolid (cluster crystal)
phase of Figure \ref{sqom_all}.
}
\end{center}
\end{figure}
Its nature is assessed by studying its behaviour as the superfluid fraction $\rho_s$ decreases on approaching the transition to the normal cluster crystal. Figure \ref{spectrum_confronto} displays $S(k,\omega)$ at the Brillouin zone edge for $r_s=0.421$, as in Figure \ref{sqom_all}b), and for a denser system at $r_s=0.389$ where the system is still supersolid but $\rho_s$ drops from 0.30 to 0.15. When the density increases the particles progressively get more localized around lattice sites, as shown by the density profiles of Figure \ref{density_confronto}, and the system gets stiffer. Correspondingly, the high energy peak of the supersolid spectrum shifts to higher frequencies, as expected for a phonon-like excitation. The low energy mode instead loses spectral weight following the loss of superfluid fraction, and shifts to lower frequencies, reducing its bandwidth. Similar results are obtained if superfluidity is suppressed by increasing the strength ($D$) of the interaction, at fixed density. The lower branch is thus seen to be largely unrelated to the spectra of the superfluid or the cluster solid phases. It could only be related to the phonon-maxon-roton of the superfluid if the density modulation of the supersolid could vanish smoothly. This possibility is preempted, however, by the first-order phase transition at the melting density of the supersolid.
\\ \indent
The structural properties of the cluster crystal\cite{saccani}, with particles hopping between adjacent lattice sites, suggest an analogy with the Bose-Hubbard model which is clearly born by the spectral properties of both systems. While in continuous space the lattice is self-assembled, and thus features vibrational modes that are obviously absent in the BHM, the lower branch of the supersolid phase shows noteworthy similarities with the acoustic excitation present in the superfluid phase of the BHM. In particular, its behaviour for wavelengths approaching the reciprocal lattice vector $\frac{4\pi}{d\sqrt{3}}$ parallels the expected linear vanishing of the superfluid mode of the BHM.
Our calculations use a finite simulation cell, implying a minimum distance of the wavevector from $\frac{4\pi}{d\sqrt{3}}$. Within this limitation, we find that the spectrum of the cluster supersolid is gapless at the reciprocal lattice vectors, i.e. there is no roton minimum. The analogy between the supersolid cluster crystal and the BHM superfluid is further supported by the softening of the low energy branch of the supersolid observed for increasing density (and/or increasing $D$), Figure \ref{spectrum_confronto}. A higher density implies reduced hopping probability and enhanced on-site repulsion, as shown in Figure \ref{density_confronto}. In the BHM this is equivalent to lowering the $t/U$ ratio, which in turn is known to reduce the bandwidth of the superfluid acoustic mode of the superfluid phase\cite{menotti}. The phase transition between supersolid and cluster solid observed in the SD system has thus some bearing with the superfluid to Mott insulator (MI) quantum transition in the BHM. This is not a full correspondence, as finite temperature and lattice dynamics contribute to stabilize a compressible cluster solid phase with non-integer mean occupation of lattice sites, more similar to the normal liquid (NL) than the MI phase of the BHM; on the other hand, the finite temperature NL and MI phases are not fundamentally different, being connected by a crossover upon varying $t/U$\cite{capogrosso1}.

In the BHM, where the translation invariance is explicitly broken, the presence of an acoustic mode is due to long-range phase coherence, which breaks the continuous gauge symmetry. In a continuous space superfluid, the corresponding mode is second sound. This mode is not seen in the spectrum of the SD superfluid phase (Figure \ref{sqom_all}a), where its spectral weight is presumably exceedingly low. In view of the analogy between the SD supersolid and the BHM superfluid, we are led to suggest that the lower branch of the supersolid is a kind of second sound. Indeed, the presence of a second longitudinal acoustic branch in a supersolid is a common feature of several phenomenological models, which assume some degree of phase coherence and the possibility of a density modulation not commensurate to the particle number\cite{Yoo,Ye,Josserand}. In particular, Ref. \onlinecite{Yoo} characterizes this second branch in the supersolid phase as a Brillouin peak, with out-of-phase fluctuations of the normal and the superfluid density, emerging from the defects-associated Rayleigh mode of the normal solid, much as second sound in superfluid $^4$He\cite{Greytak}. 
\\ \indent
Summarizing, we have carried out a numerical study of the excitation spectrum of a model cluster supersolid. The main finding is that two well-defined, distinct acoustic modes are present in the supersolid phase. The higher-energy branch is determined by the lattice dynamics, while the softer mode is uniquely due to the presence of a finite superfluid fraction. Its dispersion closely parallels that of the excitation spectrum of a superfluid Bose-Hubbard model: it further softens as superfluidity is demoted approaching the insulating solid phase; furthermore, for the system sizes studied here, it looks linearly vanishing at the reciprocal lattice vectors, rather than featuring a finite roton minimum. The experimental realization of a SS system similar to that studied here appears possible in assemblies of ultracold atoms\cite{cinti}. The double acoustic excitations, peculiar to the supersolid phase, could be detected via Bragg Spectroscopy\cite{Stamper-Kurn,inguscio}. This could be an interesting experimental verification of our findings, and a tool for identifying the supersolid phase unambiguously as well.

\section{Aknowledgements}
We thank the authors of ref. \cite{Vitali} for providing us with the GIFT code and Dr. E. Vitali for assistance with the GIFT inversions. This work was supported by the Natural Science and Engineering Research Council of Canada under research grant G121210893.
 

\begin{thebibliography}{99}



\bibitem{Gross}
Gross, E. P. Unified Theory of Interacting Bosons.
{\it Phys. Rev}. {\bf 106}, 161 (1957).

\bibitem{Andreev_Lifshitz}
Andreev, A. F. \& Lifshitz, I. M. Quantum theory of defects in crystals.
{\it Sov. Phys. JETP} {\bf 29}, 1107 (1969).

\bibitem{Thouless}
Thouless, D. J. The flow of a dense superfluid.
{\it Ann. Phys.} {\bf 52}, 403 (1969).

\bibitem {Leggett}
Leggett, A. Can a solid be superfluid?
{\it Phys. Rev. Lett.} {\bf 25}, 1543 (1970).

\bibitem{Chester}
Chester, G. V. Speculations on Bose-Einstein Condensation and Quantum Crystals.
{\it Phys. Rev. A} {\bf 2}, 256 (1970).

\bibitem{rmp}
See, for instance, Boninsegni, M. \& Prokof'ev N. V. Supersolids: What and Where are they ?
ArXiv:1201.2227, to appear in Rev. Mod. Phys.



\bibitem{cinti}
Cinti, F. {\it et al.} Supersolid Droplet Crystal in a Dipole-Blockaded Gas.
{\it Phys. Rev. Lett.} {\bf 105}, 135301 (2010). 


\bibitem{saccani}
Saccani, S., Moroni, S. \& Boninsegni, M. Phase diagram of soft-core bosons in two dimensions.
{\it Phys. Rev. B} {\bf 83}, 092506 (2011). 

\bibitem{Likos1}
Likos, C. N., Lang, A., Watzlawek, M. \& L{\"o}wen, H. Criterion for determining clustering versus reentrant melting behaviour for bounded interaction potentials.
{\it Phys. Rev. E} {\bf 63}, 031206 (2001).

\bibitem{lukin}
Lukin, M. {\it et al.} Dipole Blockade and Quantum Information Processing in Mesoscopic Atomic Ensembles.
{\it Phys. Rev. Lett.} {\bf 87}, 037901 (2001).

\bibitem{Stamper-Kurn}
Stamper-Kurn, D. M. {\it et al.} Excitation of Phonons in a Bose-Einstein Condensate by Light Scattering.
{\it Phys. Rev. Lett.} {\bf 83}, 2876 (1999).


\bibitem{inguscio}
Cl\'ement, D., Fabbri, N., Fallani, L., Fort, C. \& Inguscio, M. Bragg Spectroscopy of Strongly Correlated Bosons in Optical Lattices.
{\it J. Low Temp. Phys.} {\bf 158}, 5 (2010).

\bibitem{sqom}
Boninsegni, M. \& Ceperley, D. M. Density fluctuations in liquid $^4$He. Path integrals and maximum entropy.
{\it J. Low Temp. Phys.} {\bf 104}, 339 (1996).

\bibitem{capogrosso}
Capogrosso-Sansone, B., Prokof’ev, N. \& Svistunov, B. Phase diagram and thermodynamics of the three-dimensional Bose-Hubbard model.
{\it Phys. Rev. B} {\bf 75}, 134302 (2007)

\bibitem{menotti}
Menotti, C. \& Trivedi, N. Spectral weight redistribution in strongly correlated bosons in optical lattices.
{\it Phys. Rev. B} {\bf 77}, 235120 (2008).

\bibitem{Yoo}
Yoo, C. D. \& Dorsey, A. T. Hydrodynamic theory of supersolids: Variational principle, effective Lagrangian, and density-density correlation function.
{\it Phys. Rev. B} {\bf 81}, 134518 (2010).

\bibitem{note}
Specifically, all the results reported here are obtained for $T=0.5$ and $D=3$, with numbers of particles up to around 500. 

\bibitem{worm} Boninsegni, M. N., Prokof’ev, N. \& Svistunov, B. Worm Algorithm for Continuous-Space Path Integral Monte Carlo Simulations.
{\it Phys. Rev. Lett.} {\bf 96}, 070601 (2006).

\bibitem{worm2} Boninsegni, M. N., Prokof’ev, N. \& Svistunov, B. Worm algorithm and diagrammatic Monte Carlo: A new approach to continuous-space path
integral Monte Carlo simulations.
{\it Phys. Rev. E} {\bf 74}, 036701 (2006).

\bibitem {Vitali}
Vitali, E., Rossi, M., Reatto, L., \& Galli, D. E. Ab initio low-energy dynamics of superfluid and solid $^4$He.
{\it Phys. Rev. B} {\bf 82}, 174510 (2010).

\bibitem{maxent}
Gubernatis, J. E. \& Jarrell, M. Bayesian inference and the analytic continuation of imaginary-time quantum Monte Carlo data.
{\it Phys. Rep.} {\bf 269}, 133 (1996). 

\bibitem{Neuhaus}
Neuhaus, T. \& Likos, C. N. Phonon dispersions of cluster crystals.
{\it J. Phys.: Condens. Matter} {\bf 23} 234112 (2011).

\bibitem{capogrosso1}
Capogrosso-Sansone, B., S\"{o}yler, \c{S}, G., B. Prokof’ev, N. \& Svistunov, B. Monte Carlo study of the two-dimensional Bose-Hubbard model.
{\it Phys. Rev. A} {\bf 77}, 015602 (2008).

\bibitem{Ye}
Ye, J. Elementary excitation in a supersolid.
{\it Europhys. Letters} {\bf 82}, 16001 (2008).

\bibitem{Josserand}
Josserand, C., Pomeau, Y., \& Rica, S. Coexistence of Ordinary Elasticity and Superfluidity in a Model of a Defect-Free Supersolid.
{\it Phys. Rev. Lett.} {\bf 98}, 195301 (2007).

\bibitem{Greytak} 
Tarvin, J. A., Vidal, F. \& Greytak, T.J. Measurements of the dynamic structure factor near the lambda temperature in liquid helium.
{\it Phys. Rev. B} {\bf 15}, 4193 (1977).


\end{thebibliography}
\end{document}